# Liquid and back gate coupling effect: towards biosensing with lowest detection limit.

*Sergii Pud†, Jing Li†, Volodymyr Sibiliev†, Mykhaylo Petrychuk††, Valery Kovalenko††, Andreas Offenhäusser†, Svetlana Vitusevich*†*

† Peter Grünberg Institute, Forschungszentrum Jülich, Jülich 52425, Germany

†† Radiophysics Faculty, Taras Shevchenko National University Kyiv, 03022 Kyiv, Ukraine

e-mail: s.vitusevich@fz-juelich.de

ABSTRACT. We employ noise spectroscopy and transconductance measurements to establish the optimal regimes of operation for our fabricated silicon nanowire field-effect transistors (Si NW FETs) sensors. A strong coupling between the liquid gate and back gate (the substrate) has been revealed and used for optimisation of signal-to-noise ratio in sub-threshold as well as above-threshold regimes. Increasing the sensitivity of Si NW FET sensors above the detection limit has been predicted and proven by direct experimental measurements.

---

* -To whom correspondence should be addressed. E-mail: s.vitusevich@fz-juelich.de

In the 21st century, the tremendous progress in understanding biological objects at a molecular level has demanded development of highly-sensitive, stable and reliable sensors[1–4]. In this respect, using field-effect transistors (FETs) as biosensors is a very promising approach, because of their ability to directly translate changes in the surface potential into modulation of a drain current[5]. A new generation of nanoscaled FET sensors, based on nanowires, nanotubes[2], nanoribbons[6] and graphene sheets[7], has emerged to fulfill the need for increased sensitivity and higher spatial resolution[8,9]. Nanoscale transducers provide enhanced surface-to-volume ratio, which results in improved sensitivity[10] to surface changes[11]. Among other devices, silicon nanowire (Si NW) FET biosensors provide a variety of shapes and sizes, as well as the reproducibility offered by silicon processing technology.[12,13] The Si NW FETs are promising for different kinds of applications, such as label-free detection of biological species [14–17] and extracellular investigation of electrogenic cell activity[18–20].

The important milestone of the research in the area of biosensors is the signal-to-noise ratio (SNR), which determines the detection limit of a given type of sensors [6,21]. Biosensors often work with detection of discrete events, such as binding of the biological objects to the surface of the nanowire or action potentials of an electrogenic cell. Therefore, in a wide range of applications, Si NW FET biosensors have to be capable of real-time data acquisition. In this case, averaging techniques, which usually operate with mean values, are not applicable, and the SNR of the raw output from the sensor should be higher than one. Implementing such strong requirements might be a rather challenging task while working with extremely small signals, such as monitoring the changes of surface potential at the level of single-molecule detection or extracellular recordings from a single cell. Therefore, understanding the factors influencing the SNR is of extreme importance for development of reliable biosensors with sufficient sensitivity. The SNR of a Si NW FET sensor is determined by the transconductance, $g_m$, of the Si NW transistor, as well as by intrinsic noise of the device. The SNR can be improved both by optimisation of the fabrication technology [22,23], to obtain low-noise devices with optimal geometry, and by finding the appropriate transport regimes to maximize the sensitivity. It has also been shown that the main sources of noise are determined by the intrinsic fluctuation processes of the FET, and not by the interface with liquid[24]. To date there remain important challenges in establishing the optimal transistor operation modes for biosensing. Currently, two main concepts are considered for optimization of sensitivity: using the subthreshold mode[25] or above-threshold mode[24].

Improving the detection limits of the fabricated FET biosensor devices by optimizing the operation regime is usually performed after all of the fabrication steps. Such an optimisation can be performed for sensor development using additional gate electrodes. It should be noted that the back gate's ability to decrease operation voltages of the device was discussed in the case of silicon nanoribbons with floating liquid gate[26]. At the same time, the back gate can be utilised to reduce the impact of contact noise on the silicon nanoribbon sensor behaviour[5] So far, determining the optimal transport regimes (back gate voltage, subthreshold and overthreshold modes of operation, and linear versus saturation mode) for achieving high SNR in biosensing Si FETs has not yet been considered.

In this paper, we address the influence of the FET biosensor's operation regime on the SNR and, in particular, we investigate channel transport at different biasing conditions. In spite of published experiments, we consider both subthreshold and overthreshold modes of transistor operation. We utilise noise spectroscopy - a powerful approach for investigating device performance and structure - for the evaluation of SNR values and characterisation of Si NW transport properties. Analysis of experimentally obtained data using existing models of noise behaviour has shown switching of scattering mechanism under influence of back gate biasing. We have shown significant improvement of the SNR by utilisation of the front-back-gate coupling effect.

The structures studied were silicon nanowire array FETs (Figure 1 (a)), which we have fabricated at the Helmholtz Nanoelectronic Facility of Forschungszentrum Jülich (Germany) on the basis of the SOITEC silicon-on-insulator wafer (buried oxide of 145 nm thickness, active Si layer was initially 70 nm thickness and initial boron doping of $10^{15}$ cm$^{-3}$), using a top-down approach. The Si layer was thinned down to 50 nm by dry thermal oxidation and subsequent HF etching. Each of the FET devices contained an array of 50 nanowires connected in parallel for higher robustness and increased current through the sample. The contact regions were implanted with boron ions to $10^{19}$

cm$^{-3}$. The described NW devices represent liquid-gated accumulation FETs with a p$^+$-p-p$^+$ doping profile. The channel of the NWs was protected from the electrolyte by 9 nm of thermally grown silicon oxide. The electrolyte solution served as a front-gate of the transistor (Figure 1(b)). Contacts to the Si NW FETs were protected from the electrolyte solution by SU8 photoresist. The silicon substrate served as a back-gate. Different FETs with Si NWs of different dimensions have been investigated. The length of the samples was varied from 2 to 22 µm to extract information about contact resistance. The widths of each single nanowire in an array varied between different devices from 100 to 500 nm. It should be noted that the results presented herein are for the Si NW array FET of 50 nanowires with width of 250 nm and channel length of 16µm. However, the effects revealed by our measurements of various size samples were similar across all of the samples measured.

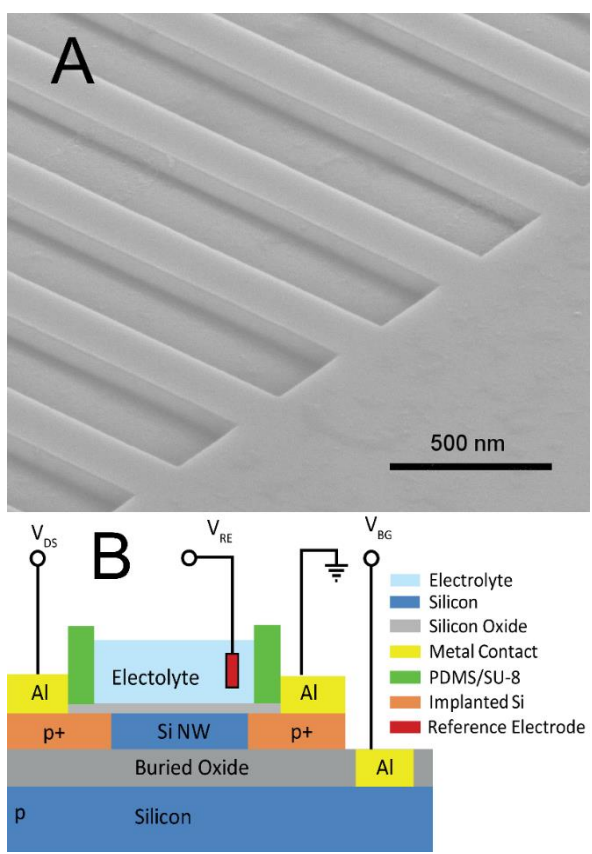

Figure 1. (a) Scanning electron microscopy (SEM) image of part of the silicon nanowire array FET sample in close vicinity of drain contact. The channel width of the single wire in array of 50 NW is 250 nm and length of each is 16 µm (b) Sketch of the experimental layout.

The schematic layout of the experiment is shown in Figure 1(b). The source of the transistor is grounded. The drain voltage, $V_{DS}$, the back-gate (substrate) voltage, $V_{BG}$, and the front-gate voltage (reference electrode), $V_{FG}$, were set against the grounded source. The drain current, $I_D$, of the FETs was measured using the method of load resistance (See Figure S1, Supporting Information). Noise spectra and time traces were acquired using an ultra-low noise preamplifier custom built for this purpose and an HP 35670 dynamic parameter analyser (Hewlett Packard).

The transconductance, $g_m$, of the FET device can be given as:

$$g_m = \frac{\partial I_D}{\partial V_{FG}} \qquad (1)$$

and can be either calculated from the transfer curve, $I_D$ – $V_{FG}$ of the FET, or measured. In this paper we report the transconductance determined by a lock-in technique (See Figure S2, Supporting Information).

All Si NW FET samples investigated can be operated by applying a voltage to the substrate, the back gate voltage ($V_{BG}$), or applying a voltage to the reference electrode, defined as the front gate voltage ($V_{FG}$). Figure 2(a) illustrates the typical $I_D$-$V_{FG}$ transfer curves measured at different $V_{BG}$. Application of the back gate potential shifts the threshold voltage of the Si NW FET. Such a behaviour is caused by the front-back gate interface coupling effect, which has been shown for metal gated FET structures[27–29]. However, this effect has not yet been reported for liquid gated Si NW FETs. Influence of the back gate potential on the threshold voltage and main NW FET sensor properties is revealed in this work. In addition, results can be used to adjust the working point of the transistor for operation with improved SNR, as will be described below.

The transconductance value of Si NW FETs determines how effectively the gate controls the drain current. As can be seen in Figure 2(a), the $g_m$, of the Si NW FET is dependent not only on front gate, but also on the back gate voltage. Therefore, the back gate voltage can tune the Si NW FET operation regime in terms of sensitivity. A summary of the transconductance measurements at different values of front and back gate voltages is shown in Figure 2(b) as contour plot. The working point of the sensor is usually chosen near the maximum value of transconductance[24]. The transconductance has a pronounced peak (Figure 2b) at certain combination of front and back gate voltage, with $V_{FG}$ of about 0 V and $V_{BG}$ of about -10V. It should be emphasised that increasing the back gate potential improves the maximum transconductance value of the Si NW FET. The increase in $g_m$ can be explained by the shift of the conducting channel from the surface to the bulk volume of the channel. There, scattering is lower than at the interface between the channel and the gate oxide, and hence, the mobility is higher.

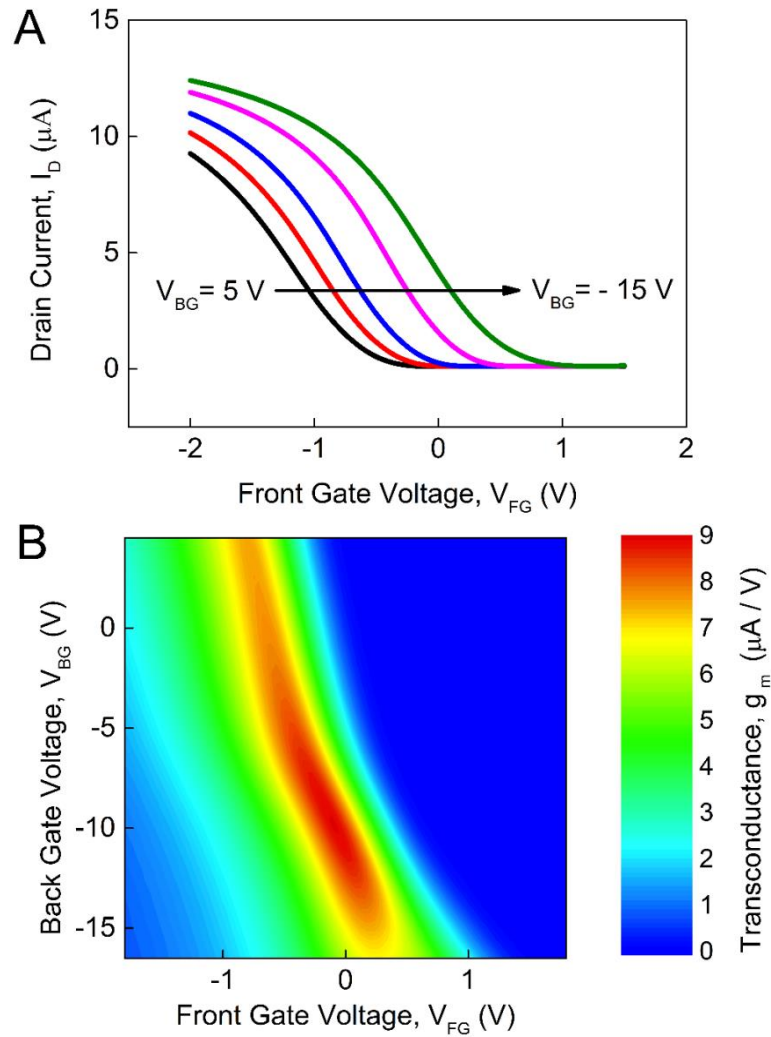

Figure 2. (a) Transfer curve of Si NW array FET with 50 nanowires (with 250 nm width and channel length of 16μm), measured at drain voltage, $V_{DS}$, of 1 V and back-gate voltages, $V_{BG}$, which varied in the range from 5V to -15V with a step size of 5V. The direction of $V_{BG}$ changes is shown with and arrow in the figure. (b) Transconductance of the Si NW array FET measured at $V_{DS}$, of 1 V and plotted in a colour map as a function of $V_{BG}$ and $V_{FG}$. Red corresponds to the maximum transconductance.

The back gate voltage is applied to the whole structure, including contact regions, as can be seen in Figure 1(b). Therefore, modulation of the transconductance by changes in the back gate voltage might be caused by changes of the contact resistance. In order to exclude any contact effects, we performed measurements of transfer curves for samples with different channel length. The experiments were carried out at small $V_{DS}$, to ensure a linear regime of operation. At small drain biases (less than the overdrive gate voltage, $V_{FG}$-$V_{Th}$) the channel of the Si NW FET sensor can be treated as a resistor, whose value is proportional to the length of the channel. Resistance of the channel measured at the same values of the overdrive voltage ($V_{FG}$-$V_{Th}$, in the linear region) should then be a linear function of the length[30]

$$R_{ch} = R_c + \frac{\rho}{S} L_{ch}, \tag{2}$$

where $R_{ch}$ is the total resistance of the Si NW channel, $R_c$ is the overall contact resistance of both source and drain regions, $\rho$ is the effective channel resistivity, $S$ is the area of the channel cross-section and $L_{ch}$ is the channel length. If we perform linear fitting of the dependence of channel resistance on length, the contact resistance can be extracted as the intercept of the fitted curve with the ordinate axis. The channel resistance measured as a function of length at a drain bias of 0.1 V and overdrive gate voltage of 0.5 V is shown in Figure 3(a) for different back gate voltages. Applying a negative (opening) back gate voltage decreases the channel resistance. The contact resistance, extracted using minimum least square fit interpolation is shown in Figure 3(b). It grows with increasing absolute value of $V_{BG}$. Taking into account that this resistance is connected in series with the channel resistance, the actual drain voltage applied to the sample is reduced. However, the transconductance increases (Figure 2b) with higher magnitude back-gate voltage. This fact reflects that increasing the transconductance is not caused by a contribution of contact regions to the transport in the Si NW array FET.

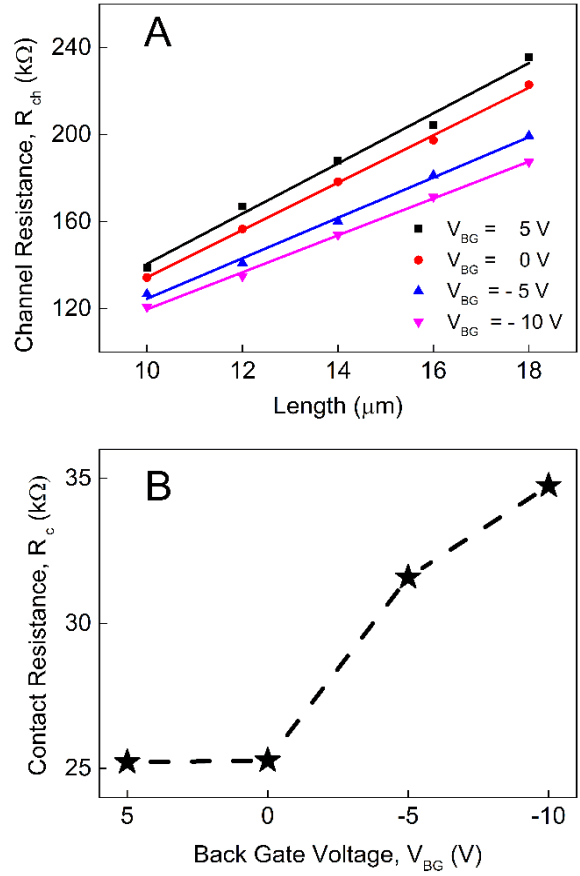

Figure 3 (a) Channel resistance of the Si NW array FET with channel width of 250 nm plotted versus channel length at different back gate voltages. The straight lines correspond to the linear least square fit of the $R_{ch}$ ($L_{ch}$) dependence. (b) Contact resistance of the Si NW array FETs plotted as a function of back gate voltage. Graph points are connected with lines as a guide to the eye

The detection capability of the FET sensor is determined by transconductance, $g_m$, as was mentioned before, and also by the intrinsic noise of the device. The change of the surface potential $\delta V_{FG}$, caused by a surface event (e.g. an action potential of a cell on top of the NW or binding of an analyte to the surface of the sensor) is registered as a change in the drain current, $\delta I$, which can be expressed as a superposition of the useful signal $\delta I_S$ and channel current fluctuations, $\delta I_{fl}$:

$$\delta I = \delta I_S + \delta I_{fl} \tag{3}$$

Signal-to-noise ratio (SNR) by its definition will then be:

$$SNR = \frac{\delta I_S}{\delta I_{fl}} \tag{4}$$

Real signals in biological environments are usually extremely small, that's why we might not be aware of the value of $\delta I_S$, because it can be below the level of channel noise. Therefore instead of

Equation (4), where we compare the response registered in the channel of the nanowire and intrinsic noise of the device, it is more suitable to use the relation between input signal $\delta V_{FG}$ and the level of equivalent input noise $\delta V_{fl}$ (the effective fluctuation of gate voltage potential, which causes corresponding fluctuation of the channel current $\delta I_{fl}$). The aforementioned relationships can be expressed as follows:

$$SNR = \frac{\delta I_S}{\delta I_{fl}} = \frac{\delta I_S\, g_m}{\delta I_{fl}\, g_m} = \frac{\delta V_{FG}}{\delta V_{fl}} \qquad (5)$$

The biosensor signals are usually detected in the low frequency range of spectra. For example, extracellular measurements of action potentials from neuronal cells[31] require detection of signals with average pulse duration about 3 ms, whose principle components are, on average, not faster than 10 kHz. Therefore the signal from sensor $\delta I$ is often filtered and restricted to the frequency range of the input signal. In this respect we should switch to the spectral representation of the equivalent input noise in Equation (5):

$$SNR = \frac{\delta V_{FG}}{\sqrt{\int_{f_1}^{f_2} S_u df}} = \delta V_{FG} \frac{g_m}{\sqrt{\int_{f_1}^{f_2} S_I df}} \qquad (6)$$

where $S_I$ is the drain current spectral density and $S_u$ is the equivalent input voltage spectral density. It is obvious that the lower value of $S_u$ leads to higher SNR. Improvement of the SNR can be realised by finding the operation modes with the lowest equivalent input noise. An approach to noise measurements in the constant resistance regime is a powerful method for finding the optimised conducting channel position[32]. In this approach, $V_{FG}$ and $V_{BG}$ are changed in such a way that drain current remains constant. The drain-source voltage $V_{DS}$ is maintained at a constant small value to ensure the device remains in the linear regime of operation. Under such conditions, the charge carriers redistribute from the top surface to the bulk of the nanowire. In the regime of constant resistance, applying a positive front-gate voltage and a negative back-gate voltage moves the conducting channel from the top interface of the NW to the bottom one, while keeping the drain current constant.

For analysis of the SNR behaviour in the regime of constant resistance, we have measured transconductance and noise of the Si NW array FET at the channel resistance of 330 kOhm, which corresponds to the maximum of $g_m$ at zero back gate voltage. Current through the sample had the value of 0.3 µA. Results of these measurements are presented in Figure 4. As one can see from Figure 4(a), the transconductance demonstrates non-monotonic behaviour as we apply negative potential to the back gate (which attracts holes to the bottom interface of the NWs) and positive potential to the front gate (which repels holes from the top surface of the NWs). We observe maximum of transconductance at the value of $V_{BG}$=-7V, which is in good agreement with what we obtain from the Figure 2 (b). Further decrease of the transconductance with increasing back gate voltage can be explained by the conducting channel shifting from the top of the dielectric layer to the bulk of the Si NW array FET. The channel current spectral density (Figure S3 of the Supporting Information) was measured in each point of this measurement. The flicker (1/f) noise component, which causes the main fluctuations of the channel current at low frequencies (1-1000Hz) is extracted from the spectra using fitting, and is shown together with the transconductance (Figure 4 (a)). The current spectral density of flicker noise follows the behaviour of the transconductance in the narrow region of back gate voltages (from 0 to -10 V). Taking into account that current was the same for the all measured points, we can conclude that the top dielectric layer is responsible for fluctuations generated in the channel of the nanowire for the case of low back gate biases.[33,34] The equivalent input spectral density value, $S_u$, calculated for the flicker noise component at 1Hz is given in Figure 4(b). In Figure 4(b) we observe a slight decrease of the $S_u$ value at more negative back gate potentials. This fact demonstrates the high quality of the buried oxide layer[35], because moving the channel from top to the bottom interface of the NW reduces $S_u$. The decrease in $S_u$ also results in the increase of the SNR (Equation (6)), which is favourable from the point of view of sensor performance. In the region of back gates between -10V and -20V the dependence of $S_u$ on back gate voltage becomes weak and also the value of $S_u$ is lower than in the case of an unbiased back gate (see Figure 4(b)). Therefore, we consider this range of back gate voltages as optimal, because the equivalent input noise is lower than in the case of zero back gate potential and scattering of the $S_u$ value for the $V_{BG}$ between -10 and -20 V is lower than for higher back gate voltages (Figure 4(b)). It should be noted that the minimum of the equivalent noise value does not coincide with the position of maximum transconductance.

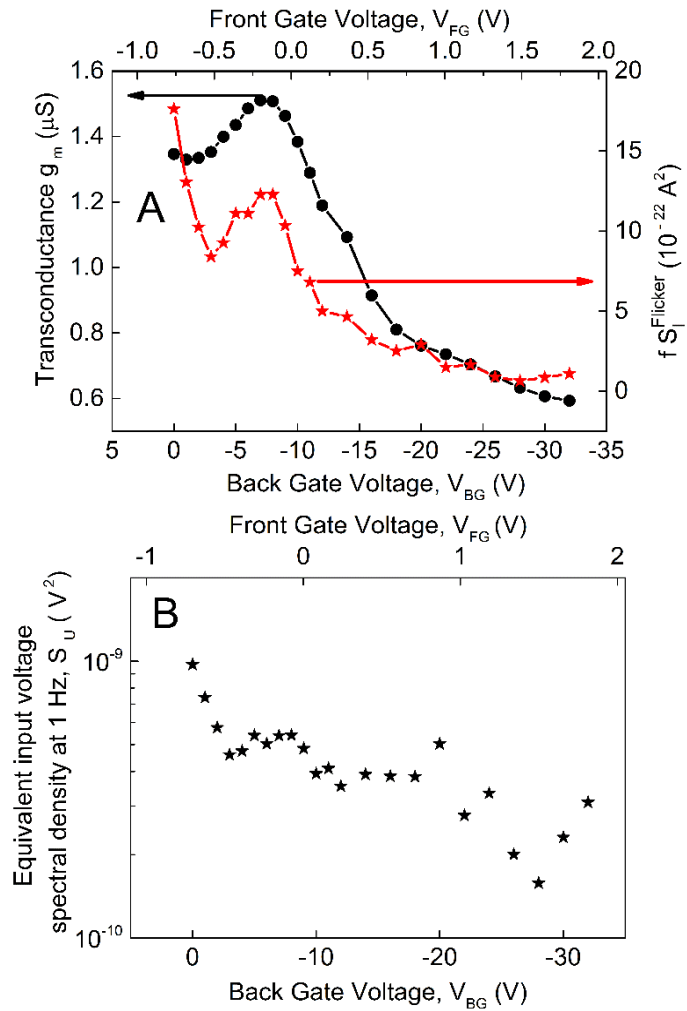

Figure 4. (a) Typical transconductance (black circles), and amplitude of flicker noise at 1 Hz (red stars), of the Si NW array FET measured in the regime of constant channel resistance. The dependencies are functions of both front (top axis) and back (bottom axis) gate voltages. The data points are connected with lines for better visual perception. (b) Equivalent input voltage spectral density calculated for the flicker noise component at 1Hz and plotted as a function of back gate and front gate voltage. The estimated error of the values presented is less than 10%.

To find the optimal regime of operation for the Si NW FET array, we have performed a calculation of signal-to-noise ratio at different drain, front-gate and back gate biases. Biosensor signals, including the action potentials of neuronal cells and label free detection of biomolecules, belong to the low frequency sensing range. Hence, a frequency range between 1Hz and 10 kHz was chosen for evaluation of the SNR to fit most bio-applications. Within this range, both excess noise and thermal noise have to be considered (Figure S3 of Supporting Information). Therefore, for calculation of the SNR we use the full spectra of $S_u$ and perform numeric integration according to

Equation (6) to get the contribution of all types of noise. The $\delta V_{FG}$ in Equation (6) is considered to be 1V.

For our measurements, we have chosen parameters that cover entire range of transport regimes. The front gate voltage was swept in the full range from the subthreshold to the overthreshold region; the main characteristic workpoints were defined as the combination of two drain and two back gate voltages. The two drain bias values were taken as 0.1 V and 1.0 V, which represent linear and saturation modes, respectively. For the back gate voltage, the following two values are used: 0 V, representing the normal case with an unbiased substrate and -10 V, which was chosen as an optimal value, using data from Figure 4(b). In Figure 5(a), the transconductance of the Si NW array FET, measured using a lock-in technique, is shown as a function of overdrive front gate voltage for the main cases described above: $V_{DS}$=0.1V, $V_{BG}$=0V; $V_{DS}$=0.1V, $V_{BG}$=-10V; $V_{DS}$=1.0V, $V_{BG}$=0V and $V_{DS}$=1.0V, $V_{BG}$=-10V. The normalised signal-to-noise ratio calculated for the measured transconductance values is shown in Figure 5(b).

At low drain biases (Figure 5(b)), the SNR follows behaviour of $g_m$ (Figure 5(a)). Such a behaviour of the normalised SNR is mainly related to the contribution of the thermal noise component within the chosen frequency range (see Figure S4 of the Supporting Information). Applying a back gate voltage has no strong impact on $g_m$ at low drain bias (Figure 5(a)), but it decreases overall channel resistance (Figure 3(a)) and hence decreases thermal noise. This results in slight increase of the SNR (Figure 5(b)) at the point of maximum of transconductance.

At the relatively high drain voltage of 1V, the maximum in the SNR dependence on front gate potential is not resolved in contrast to the linear region behaviour ($V_{DS}$=0.1V). It reflects the major contribution of the flicker and generation-recombination noise components (Figure S4 of Supporting Information), which change proportionally to the $g_m$. Therefore the integrated value of $S_u$ has a flat region at the front gate voltages, where the transconductance reaches its maximum. The normalised SNR at 1V drain voltage (Figure 5(b)) is much higher than one at 0.1V. At low drain voltage, the drain current and the transconductance of the sample are proportional to the drain-source voltage. By further increasing the $V_{DS}$, the drain current reaches the saturation value. This saturation regime can be used for detection of small signals, because in this mode, the drain voltage has almost no effect on drain current or the value of the flicker noise and the transconductance value is always proportional to the drain voltage.

As follows from the data in Figure 5(b), applying the back gate voltage of -10V improves SNR of the transistor up to 1.5 times when operating in saturation mode. At the same time, the transconductance value increases by approximately 1.1 times at the maximum point. It demonstrates that applying the back gate voltage reduces the noise of the device channel and therefore, the integrated value of $S_I$. It should be emphasised that the back gate voltage has a positive impact on SNR for both subthreshold and overthreshold modes (Figure 5(b)). This fact was not previously reported in literature.

Described effect of the back gate voltage on sensitivity can be explained analysing noise behavior. In Figure 5(c) we have plotted equivalent input noise spectral density extracted for 1/f noise from the spectra at 1Hz for the case of saturation mode ($V_{DS}$=1V) and different back gate voltages (0V and -10V). As it can be seen $S_U$ demonstrated weak dependence on overdrive gate voltage at $V_{BG}$=0 contrary to the case of $V_{BG}$= -10V, when $S_U$ changes proportionally to the gate voltage. The $S_u$ independence of gate voltage is the sign for number fluctuation noise model[36,37], whereas proportionality to the gate voltage reflects mobility fluctuations noise model so called volume noise[36,37]. Number fluctuations are caused by interactions of charge carriers of the channel with traps located in the gate dielectric, therefore such noise is also called surface noise.

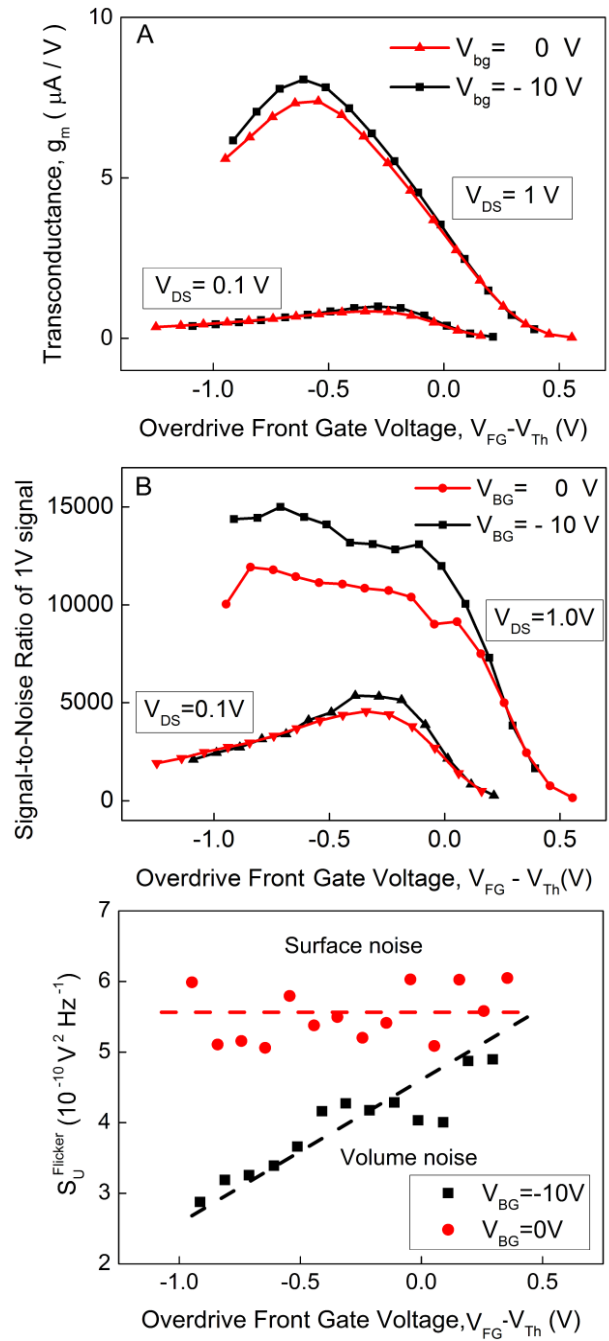

Figure 5 (a) Transconductance of the Si NW array FET; (b) signal-to-noise ratio plotted as a function of overdrive front gate voltage at two drain biases (0.1 V and 1.0 V) and different back gate voltages (0V plotted with red triangles and -10V plotted with black squares); (c) – equivalent input noise spectral density for 1/f component extracted from measured spectra and plotted versus overdrive front gate voltage at drain bias of 0.1 V and different back gate voltages, $V_{BG}$ : red circles – 0V, black squares - (-10V); red dashed line reflects $S_U$ behavior for the case of number fluctuations model of 1/f noise; back dashed line reflect proportionality to front gate voltage – typical for mobility fluctuation model .

Figure 5(c) demonstrates, that applying back gate potential switches the dominant mechanism of carrier scattering in the sample from surface to volume nature. In this case at high back gate voltages the noise suppression is caused by change in the fluctuation mechanisms (Figure 4(a)). This effect dominates over the decrease of the transconductance (Figure 4 (a)) due to change in channel position. These data represent background for increased sensitivity of silicon nanowire FETs under biased back gate conditions.

The optimised conditions obtained from the results shown in Figure 5(b) are proven by direct experimental measurements. Pulse signals with parameters similar to the extracellular recording of action potentials from a neuronal cell (pulse width of 3 ms) were applied to the reference electrode and amplitude of the pulses was adjusted to obtain a sensor response with SNR approximately one. The transistor working point was set in the following way: $V_{DS}$ was set to 2.0V (saturation mode) and $V_{FG}$ was adjusted to the value of -1.3V to get the maximum of the transconductance, $V_{BG}$ was set to 0. Figure 6(a) shows the test signal and response of the sensor. Next, the test signal amplitude was reduced by a factor of approximately 1.5, according to the results discussed previously (Figure 6(b)).

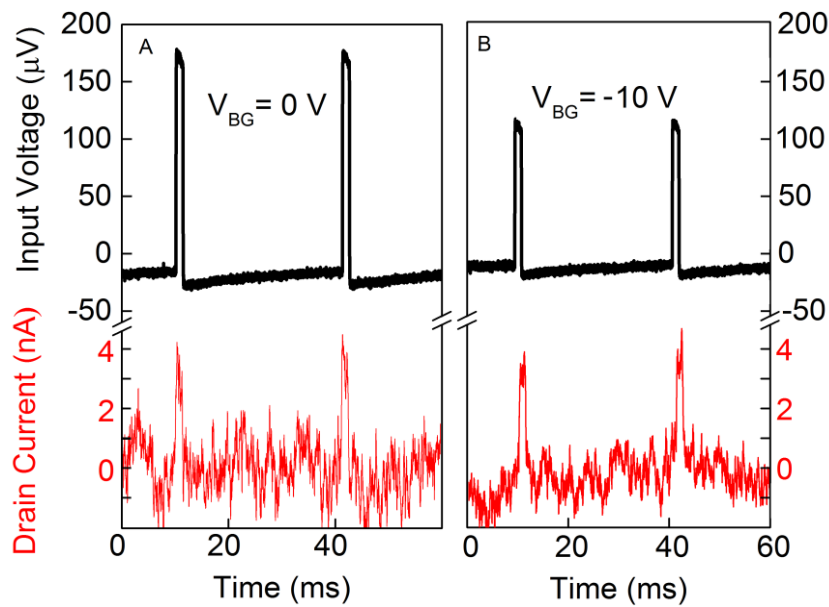

Figure 6. Experimental demonstration of the signal recovery below the device noise level, utilising front-back gate coupling. (a) Test signal (black line) applied to the reference electrode and the response of the sensor (red line) at $V_{DS}$=2.0V, $V_{FG}$=-1.3V, $V_{BG}$= 0 V. (b) Signal is well resolved at $V_{DS}$=2.0V, $V_{FG}$=-0.3V, $V_{BG}$= -10 V, with the test signal decreased by a factor of 1.5 compared to (a). Both graphs (a) and (b) contain the signals, which were filtered to the range from 1Hz to 10 kHz.

Then, using the appropriate back gate voltage (-10V) and a front gate voltage adjusted to yield the maximum transconductance, we distinguished signal despite its reduced amplitude. Figure 6(b) shows that the signal is 1.5 times lower than the initial pulse is resolved with even better SNR than it is in Figure 6(a). Hence, utilisation of the back gate electrode allows improvement of the sensitivity to signals that are otherwise below the noise limit of the sample when no back gate voltage is applied.

In conclusion, high-quality silicon nanowire array FET biosensors with different channel dimensions were fabricated and studied. An effect of coupling between the liquid gate and the substrate (back gate) has been revealed. Noise behavior in regime of biased back gate voltage is explained in frame of mobility fluctuations model whereas in the case of unbiased back gate the Si NW conducting channel behavior corresponds to number fluctuations one. Reassembling of the channel carriers changes the dominant scattering mechanism in the FET. The effect is used for tuning of the FET operation mode to increase the sensitivity of the fabricated biosensor structure by 50 %. The results reflect novel ways for improving sensitivity of the biosensors by designing devices with controlled channel position. The developed approach is used to recover signal below the initial detection limit.

ACKNOWLEDGMENT

S.Pud greatly appreciates a research grant from the German Academic Exchange Service (DAAD).

J.Li would like to acknowledge support from the China Scholarship Council.

Authors are grateful to Dr. Vanessa Maybeck for valuable discussions.

ABBREVIATIONS

Si NW, silicon nanowire; FET, field-effect transistor; MOSFET metal oxide semiconductor FET; SNR, signal-to-noise ratio;

SUPPORTING INFORMATION

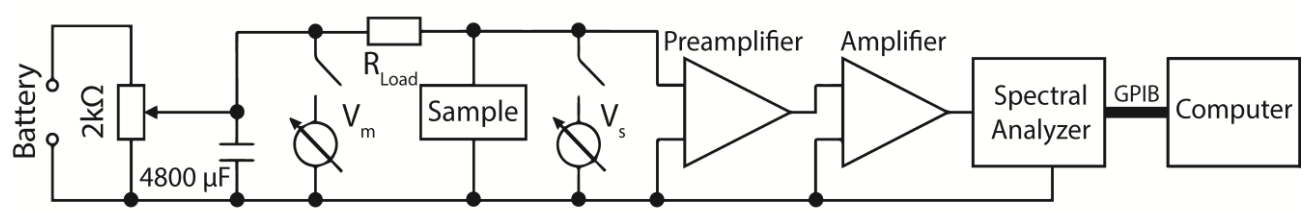

Figure S1. Schematic image of the noise measurement setup. The voltage is applied to the sample using a battery loaded on the variable resistor. Current through the sample is calculated using the difference between readings of the voltmeters $V_m$ and $V_s$. Spectra are aquired using the dynamic signal analyzer HP 35670 (Hewlett Packard).

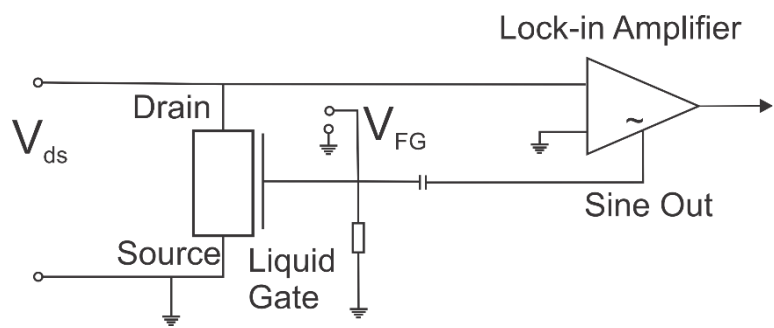

Figure S2. Schematic image of the lock-in technique for measuring the transconductance, $g_m$. The small AC sine signal (30mV) is passed through the high pass filter from the lock-in amplifier generator to the reference electrode, which serves as a liquid gate. A high pass filter is used to prevent the DC voltage $V_{FG}$ from biasing the sine generator. Then, the lock-in amplifier records the fluctuations of the drain voltage caused by the voltage applied to the sample using a battery loaded on the variable resistor.

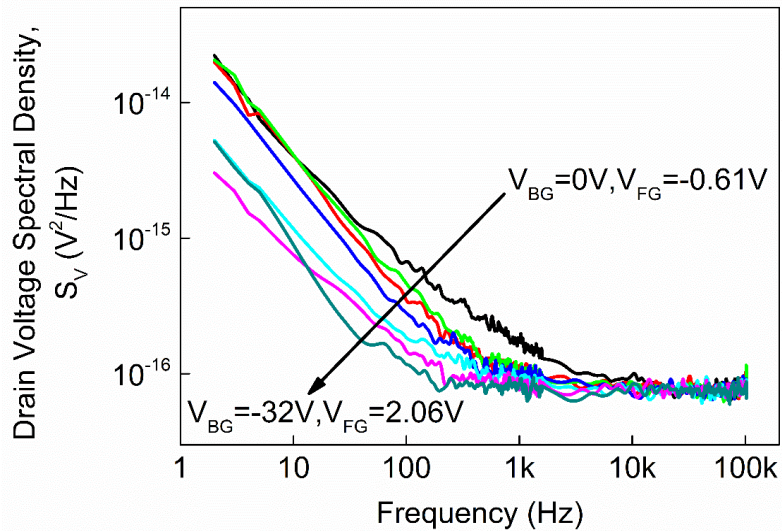

Figure S3. Drain voltage noise spectral density, $S_V$, of the array of 50 Si NWs with 250 nm width and 16 μm length, measured in the regime of constant channel resistance. Drain bias was 0.1V, front and back gate voltages were swept in such a way that the current through the sample remained constant at 0.3 μA. The direction of the gate voltage change is shown by an arrow on the graph.

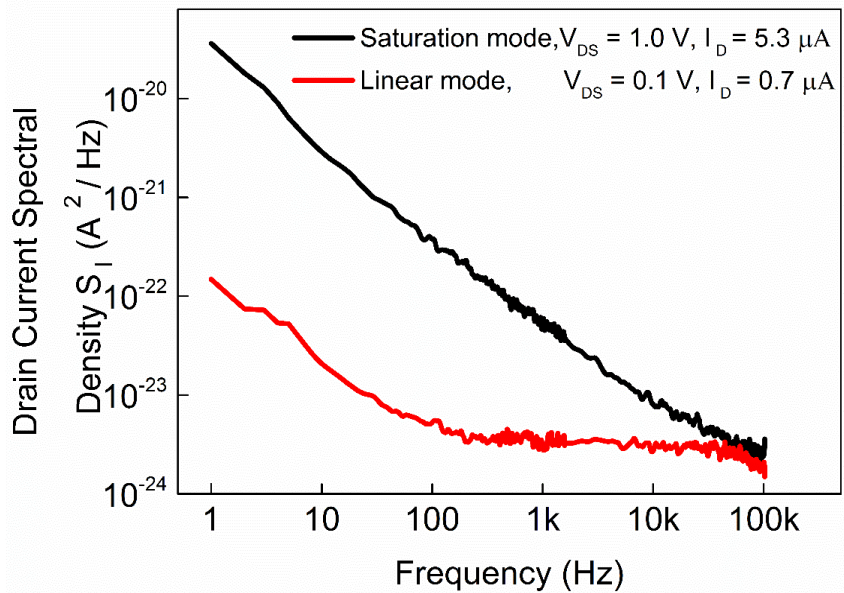

Figure S4. Comparison of the current noise spectral density for the cases of linear ($V_{DS}$=0.1V, red curve) and saturation ($V_{DS}$=1.0V, black curve) modes at an overdrive gate voltage of -0.7 V.

TOC Image

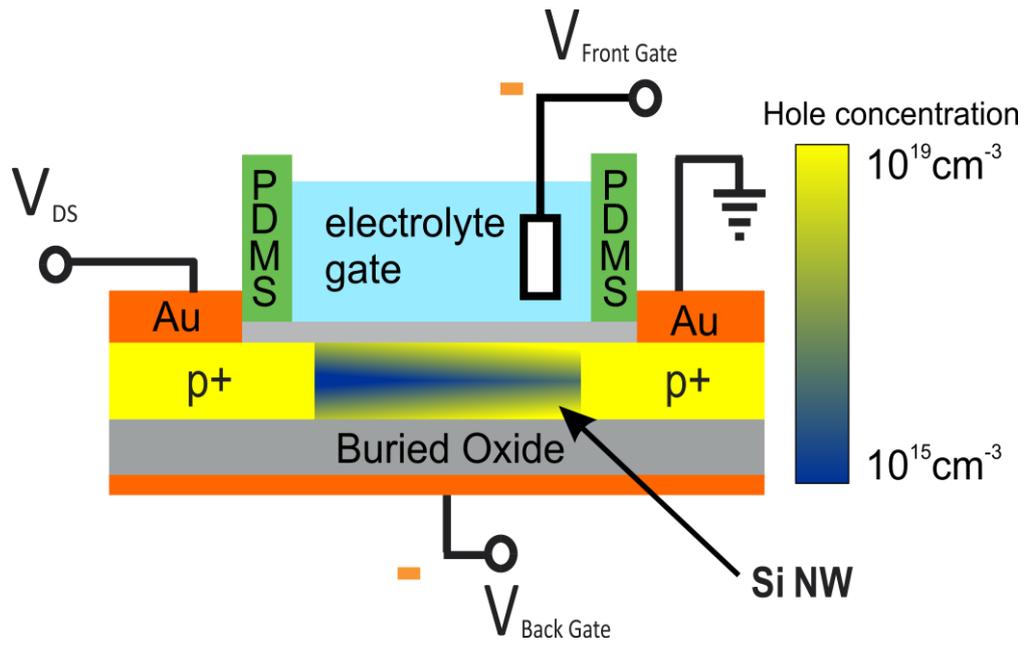